\documentclass[superscriptaddress,
reprint,
 amsmath,amssymb,
 aps,
prl,
]{revtex4-2}

\usepackage{graphicx}% Include figure files
\usepackage{dcolumn}% Align table columns on decimal point
\usepackage{bm}% bold math
\usepackage{hyperref}% add hypertext capabilities
\usepackage{lineno}
\usepackage{color,soul}
\usepackage{soul}
\sethlcolor{yellow} % Set highlight color, highlight with command \hl{}
\begin{document}

\preprint{APS/123-QED}

\title{Electron beam characterization via fluorescence imaging of Rydberg states in atomic vapor} % Force line breaks with \\

\author{Rob Behary}
\email{rbehary@wm.edu}
\affiliation{Department of Physics, William \& Mary, 300 Ukrop Way, Williamsburg, VA 23185, USA}

\author{Kevin Su}
\affiliation{Department of Physics, William \& Mary, 300 Ukrop Way, Williamsburg, VA 23185, USA}
\author{Nicolas DeStefano}
\affiliation{Department of Physics, William \& Mary, 300 Ukrop Way, Williamsburg, VA 23185, USA}
\author{Mykhailo Vorobiov}
\affiliation{Department of Physics, William \& Mary, 300 Ukrop Way, Williamsburg, VA 23185, USA}
\author{T. Averett}
\affiliation{Department of Physics, William \& Mary, 300 Ukrop Way, Williamsburg, VA 23185, USA}
\author{Alexandre Camsonne}
\affiliation{Thomas Jefferson National Accelerator Facility, 12000 Jefferson Avenue, Newport News, VA 23606, USA}
\author{Shukui Zhang}
\affiliation{Thomas Jefferson National Accelerator Facility, 12000 Jefferson Avenue, Newport News, VA 23606, USA}
\author{Charlie Fancher}
\affiliation{The MITRE Corporation, McLean, VA 22102, USA}
\author{Neel Malvania}
\affiliation{The MITRE Corporation, McLean, VA 22102, USA}
\author{Eugeniy Mikhailov}
\affiliation{Department of Physics, William \& Mary, 300 Ukrop Way, Williamsburg, VA 23185, USA}
\author{Seth Aubin}
\affiliation{Department of Physics, William \& Mary, 300 Ukrop Way, Williamsburg, VA 23185, USA}
\author{Irina Novikova}
\affiliation{Department of Physics, William \& Mary, 300 Ukrop Way, Williamsburg, VA 23185, USA}

\date{\today}% It is always \today, today,

\begin{abstract}
We demonstrate an all-optical, minimally invasive method for electron beam ($e$-beam) characterization using Rydberg electrometry. The $e$-beam passes through a dilute Rb vapor prepared in a quantum superposition of ground and Rydberg states that reduces resonant absorption in a narrow spectral region.
Imaging the modifications of Rb fluorescence due to shifts in the Rydberg state from the $e$-beam electric field allows us to reconstruct $e$-beam width, centroid position, and current. 
We experimentally demonstrate this technique using a 20~keV $e$-beam in the range of currents down to ~20 $\mu$A, and discuss technical challenges produced by environmental electric potentials in the detection chamber. Overall, we demonstrate the promising potential of such an approach as a minimally invasive diagnostic for charged particle beams.
%Through this technique we show that we can measure a 20 keV electron beam down to ~20 $\mu$A and discuss problems with our rubidium chamber that can be upgraded to improve sensitivity.
%\begin{description}
%\end{description}
\end{abstract}

\maketitle

%\tableofcontents

%\section{Introduction}
%\textit{Beginning - brief motivation for our research from accelerator physics point of view (Shukui)} 
With the steady technological advancements at high energy particle accelerators, there has been an increasing demand for more robust non-invasive spatial beam property diagnostics.
While different approaches and apparatuses exist, optical and imaging methods based on signals such as fluorescence~\cite{ImpactFluorescence,Sandoval:1994noe}, synchrotron radiation~\cite{Synchrotron}, and X-rays generated by moving particles~\cite{OTR} have played an essential role in accelerator research and beam operations.
However, each of these beam diagnostics also bear some intrinsic drawbacks and their applicability is often limited by factors such as sensitivity and system complexity. 
For example, synchrotron radiation only exists near particle trajectory-bending components~\cite{Synchrotron,LHC_Syncrotron}, Compton scattering laser wire requires high laser intensity and slow scanning between particle and laser beams~\cite{Laser_Compton_Scattering,blair2008laser}, and the gas-ionization-based 2D super-sonic gas curtain devices rely on complex mechanical systems and have low sensitivity~\cite{PhysRevAccelBeams.20.062801, PhysRevAccelBeams.27.043201}.
The work presented in this letter describes our effort in the development of a sensitive, minimally invasive apparatus capable of measuring multiple charged particle beam parameters including width, centroid position, and current simultaneously.

\begin{figure*}[ht]
\centering
\includegraphics[width=0.9\linewidth]{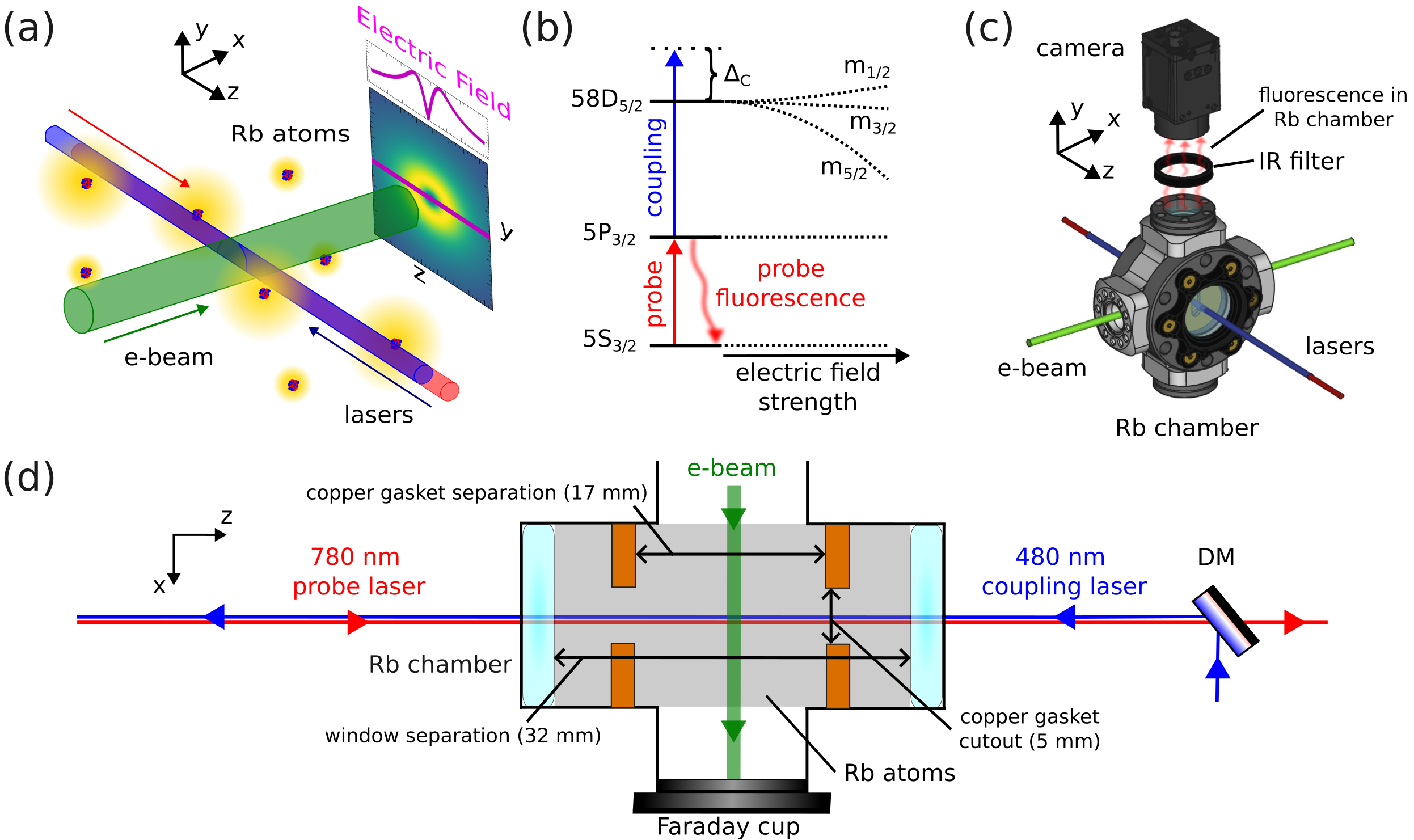}
\caption{Overview of experimental design.
(a) A charged particle beam produces an electric field and passes through a cloud of Rb atoms shown as a nucleus with a yellow electron cloud.
%\red{I first though the `Rb atoms' label in 1a) was associated with the red laser arrow.  Perhaps add a little arrow pointing to one of the atoms,} 
We use lasers to excite these atoms to a Rydberg state (shown as atoms with a larger electron cloud) to probe the electric field. We assume the beam produces a radially symmetric Gaussian electric field distribution described by Eq.(\ref{eq:e_field_e_beam}).
(b) Energy level diagram for Rydberg EIT. In the presence of an electric field, the $|m_j|$ sublevels of the Rydberg state will shift differently based on the strength of electric field.
(c) Experimental setup to capture the probe laser fluorescence in the Rb chamber with a camera.
(d) Cross section of Rb chamber on our beam line for electric field detection. Distances are marked to give a sense of scale of the setup.}
\label{fig:setup}
\end{figure*}

In this letter we apply recently developed quantum electrometry~\cite{Gallagher_1994, Nik_Rydberg_Review,SIMONS2021100273,FrancherIEEE2021} for detection and diagnostic of a charged particle beam. We take advantage of the exceptionally large electron dipole moment of Rydberg states of alkali metal atoms ($n \ge 20$) that have already been explored for a wide range of applications.
These include  SI-traceable electric field standards~\cite{HollowayJAP2017}, rf-field receivers~\cite{Nik_Rydberg_Review,KermitRFReciever,FrancherIEEE2021}, THz-imaging~\cite{Simons:18,Downes_2023,PhysRevX.10.011027}, magnetic field sensing~\cite{RydbergMagnetic,Noah2DFields}, and thermometry of blackbody radiation~\cite{RydbergBlackbody}.
Here we introduce a diagnostic method of an electron beam ($e$-beam) via all-optical measurement of its effect on the Rydberg states in dilute rubidium (Rb) vapor.
%Here we introduce a dilute Rb vapor near the electron beam, and use an all-optical method for measuring electric field produced by the electron beam. 
The reconstruction of electric field, created by the electrons, lets us calculate the centroid position, width and total current of the $e$-beam. Unlike the  complimentary technique based on magnetic field reconstruction using nonlinear magneto-polarization rotation~\cite{DeStefano_APL2024}, the current method offers superior spatial resolution, and in principle allows for three-dimensional reconstruction of the charged particle distribution.  

For all-optical detection we use  a coherent two-photon process known as electromagnetically induced transparency (EIT)~\cite{Scully,Finkelstein_2023}. The Rb atoms are excited to a Rydberg state using two laser fields: a near-infrared probe field, resonant with the $5S_{1/2}\rightarrow 5P_{3/2}$ optical transition (wavelength 780~nm), and a 480~nm blue control field that couples the intermediate $5P_{3/2}$ state with the desired Rydberg state (in our experiment $58D_{5/2}$). In this configuration the population of the intermediate state diminishes when the sum of the laser frequencies matches the frequency difference between the ground and Rydberg states, and this change can be detected either through increased probe laser transmission, or reduced infra-red (IR) fluorescence~\cite{Keaveney_2014,lukema} at the two-photon resonance condition.
Such two-photon EIT arrangement allows for relatively simple and sensitive detection of Rydberg state frequency shifts caused by external electric and magnetic fields. 
%For room temerature rubidium vapor, experiments use a 780 nm probe laser and 480 nm coupling laser overlapped in a vapor cell ~\cite{lukema,OptimizingEITVapor,RydbergInversePopSampling,Su2024}
%to reduce Doppler broadening of the EIT
%, and changes to the probe laser absorption is monitored on a photodiode~\cite{lukema,OptimizingEITVapor,RydbergInversePopSampling,Su2024}.
To maximize the interaction volume and, more importantly, to minimize the two-photon Doppler broadening~\cite{Finkelstein_2023}, two optical fields are typically counter-propagating. Keeping track of the probe laser transmission complicates extracting any spatial information about $e$-beam, since it reflects cumulative variations in the electric field along the interaction volume.
There are some strategies for spatial mapping~\cite{MaOE20,PhysRevApplied.13.054034}, using, e.g., crossed laser beam~\cite{Su2024} or a three-photon ``star'' configuration~\cite{Three_Photon_Star}, but at expense of reduced spectral resolution and lower signal-to-noise. 
%Another solution to avoid the problem of Doppler broadening is to use a three photon EIT process in a star configuration for truly localized sensing~\cite{Three_Photon_Star}, but this limits the spatial resolution to a fixed point that is difficult to translate.

The motivation of this work is to offer an \textit{in situ} minimally invasive detection tool for charged particle beams as an alternative to more standard destructive diagnostics.
In this letter, we demonstrate the possibility to map the electric field  produced by an $e$-beam by capturing Rb fluorescence, following a recently proposed fluorescence-based imaging technique~\cite{Noah2DFields,patrick2025imaginginducedsurfacecharge}.
By recording and analyzing the spatial variation of IR fluorescence in the Rydberg EIT configuration, we map spatial variations in dc electric fields along the laser beam propagation (potentially in 2 dimensions)~\cite{patrick2025imaginginducedsurfacecharge}. Here we focus on diagnostics of the electric field produced by an $e$-beam. Despite parasitic background electric fields, we are able to infer the $e$-beam width and centroid position, as well as its current.  
%However, this method may be adopted for measuring electric field distribution inside low-density plasmas ~\cite{Mordjick_BRN2024}.
%Through this method we are able to detect the electric field produced by an electron beam.
%We show that the width of the recovered electric field signal is related to the beam size so we can determine the full width at half maximum (FWHM), and monitor the beam position within our Rb chamber.

%Our signal is tainted by a highly varying background field, so further improvements to the Rb chamber need to be implemented to mitigate charging and parasitic fields for more informative diagnostics such as precise e-beam current monitoring.

%\section{Overview of Electron Beam Apparatus and Fluorescence Based Measurements}
%\label{sec:overview}

\begin{figure}[ht]
\centering
\includegraphics[width=1.0\linewidth]{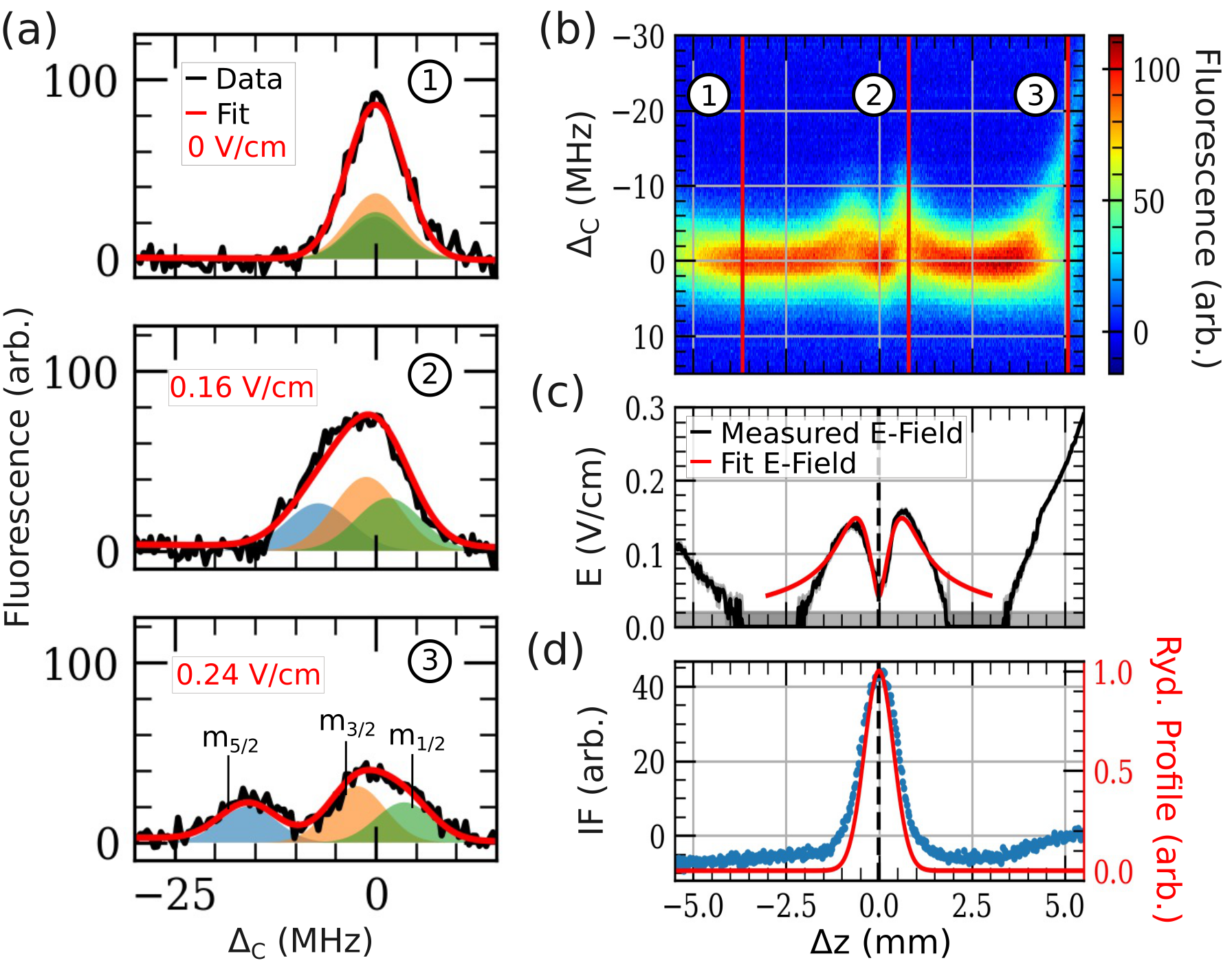}
\caption{Demonstration of fluorescence based measurements in the presence of $e$-beam.
(a) The EIT spectra that constitute the heat map in (b). The spectra are fit with Eq. (\ref{eq:sum_of_gaussians}) to determine the electric field value shown as the shaded regions in each plot.
(b) The measured spectra of the 58D${}_{5/2}$ Rydberg EIT peak for each position along the laser beam fluorescence. The numbers correspond to the shown single spectra in (a).  
(c) The reconstructed electric field value along the Rb chamber. The measured field is fit with a function described by Eq. 
(\ref{eq:e_field_e_beam}). The minimum detectable field is shown as a strip along the bottom of the plot, and error bars on the fit are shown in grey.
(d) $E$-beam profile obtained from IF. This shows a secondary detection method to verify where the $e$-beam is inside the Rb chamber.}
\label{fig:fluorescence-detection}
\end{figure}

The experimental setup and general detection principle for the $e$-beam diagnostics is shown in Fig.~\ref{fig:setup}. 
A collimated beam of 20~keV electrons passes through an area filled with a dilute Rb vapor. The Rb density of $\approx$ 1.6$\times$ 10${}^{11}$ atoms/cm${}^3$ is produced using an ampule of metallic Rb placed in the heated vacuum chamber. Fig.~\ref{fig:setup}(a) also shows a radially symmetric electric field distribution, produced by an $e$-beam with Gaussian transverse profile. Two counter-propagating laser beams, perpendicular to the $e$-beam, are used to excite Rb atoms to the 58D Rydberg state. Electric field lifts the degeneracy of this state, producing quadratic dc-Stark shifts for each of the $|m_j|$ sublevels, as shown in Fig.~\ref{fig:setup}(b).
Approximate shifts to the Rydberg sublevels are described by:
\begin{equation}
\label{eq:stark-shift}
    h \cdot \Delta f_{|m_j|}(E) \approx -\frac{1}{2}\alpha_{|m_j|}E^2.
\end{equation}
Where $h$ is Planck's constant, $\alpha_{|m_j|}$ is the scalar polarizability of the different $|m_j|$ sublevels of the Rydberg state and $E$ is the magnitude of the electric field~\cite{Noah2DFields,patrick2025imaginginducedsurfacecharge,HollowayJAP2017,Nik_Rydberg_Review}.
A more accurate description of these shifts are interpolated from a numerically solved Stark map generated by the Alkali Rydberg calculator (ARC)~\cite{ARC}.% and the relative strengths of these shifts are shown in Fig.~\ref{fig:setup} (b).

%The electric field generated by the e-beam is measured with rubidium atoms excited to a Rydberg state shown in Fig.~\ref{fig:setup} (b).
%Our e-beam is generated from a Staib instruments filament source, and  is fixed to have a Gaussian profile with a FWHM of 1.07 $\pm$ 0.06 mm and the electrons have an energy of 20 keV.

%with a total pressure in the beam line of 2 $\times 10^{-8}$ Torr.

The effect of dc Stark shifts on EIT resonance at different points along the laser beam is monitored using a CCD camera.
The camera captures the probe laser fluorescence in the Rb chamber Fig.~\ref{fig:setup}(c). For these measurements we need to capture the variation in 5P${}_{3/2}$ state population as a function of the coupling laser frequency at each point along the laser beams. To do that the camera collects images at a fixed frame rate with 60 ms exposure time for 600 frames while  the coupling laser sweeps across the 58D Rydberg state at a rate of 300~s. This way we are able to record fluorescence spectrum for each image pixel, providing localized information of electric field. We used the known frequency separation of the 58D\textsubscript{5/2} and 58D\textsubscript{3/2} Rydberg states~\cite{ARC} to calibrate the frequency axis.
For these measurements both laser fields were collimated to similar beam size [0.33~mm full width at half maximum (FWHM)]. The 780~nm probe laser has power 5~$\mu$W, and the maximum power of the 480~nm coupling laser is 70~mW. 
We used a 50~mm lens to image the plane of the laser fluorescence in the Rb chamber into the camera.
An IR filter in front of the lens eliminated background light and increased the $SNR$ for the probe laser fluorescence. The scaling of this imaging system was calibrated using a ruler.
%Fig.~\ref{fig:fluorescence-detection} (a) shows a fixed image of the laser beam fluorescence, but the coupling laser is swept across the 58D Rydberg resonance at a rate of 150 s and the camera collects images at a fixed frame rate.
%This video is changed from from time to frequency by a reference EIT cell that is monitored independently in the experimental setup.
%The fluorescence detection scheme is shown in Fig.~\ref{fig:setup} (c). 
%Frequency separation of the 58D${}_{5/2}$ and 58D${}_{3/2}$ hyperfine Rydberg states are monitored in a separate reference EIT cell and used to calibrate the frequency axis.
%, and image size is calibrated to a picture of a ruler.
%For atomic measurements, the fluorescence of the 5P${}_{3/2}$ state is monitored.

\begin{figure*}[ht]
\centering
\includegraphics[width=0.85\linewidth]{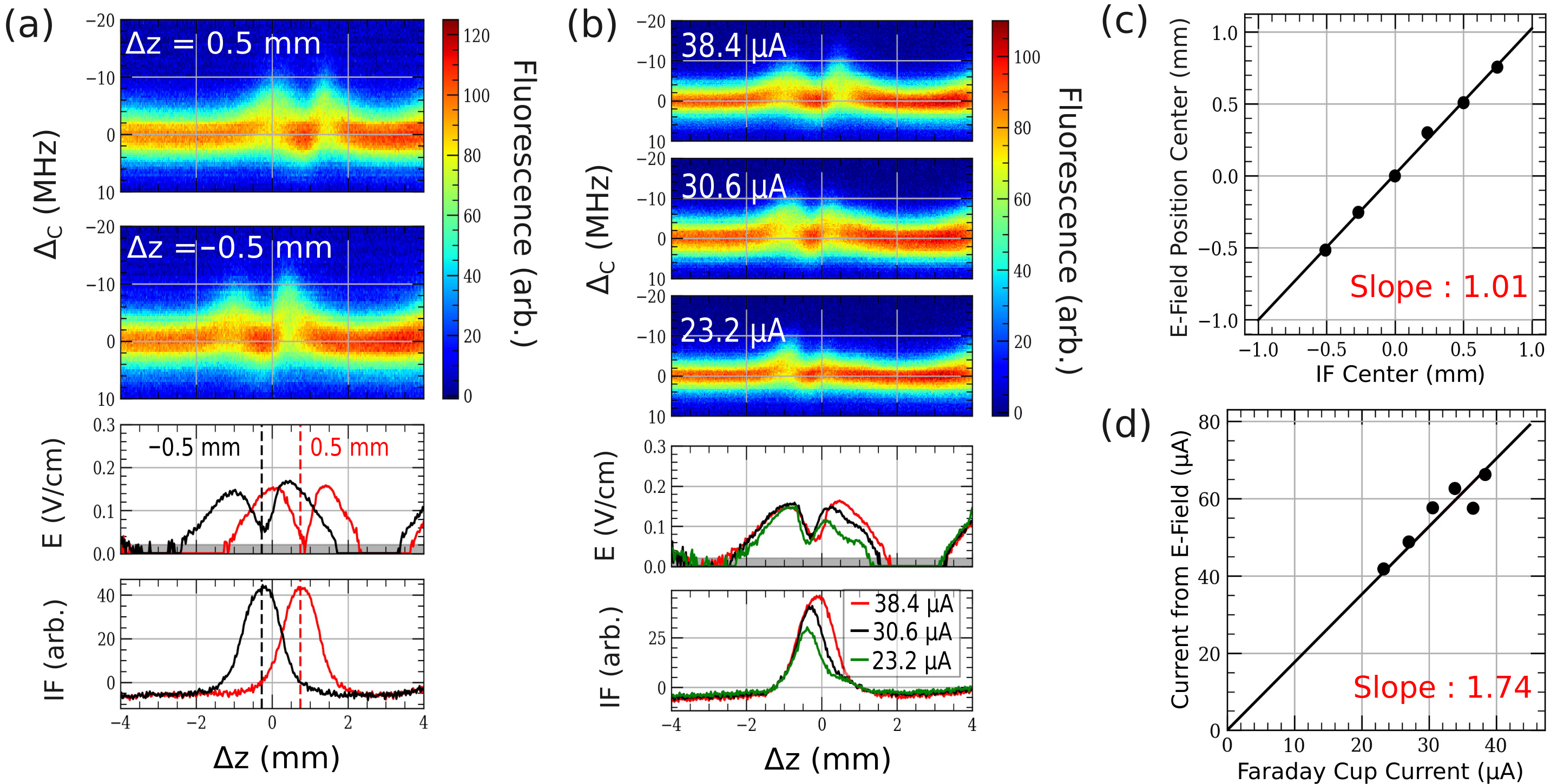}
\caption{$E$-beam diagnostics for position and current. (a) and (b) are the same style of plots shown in Fig.~\ref{fig:fluorescence-detection} (b)-(d).
(a) Heat maps for two different $e$-beam positions in the cell. The dashed lines in the plots show the beam center.
(b) Plots of heat maps for increasing $e$-beam current. Current values recorded from the Faraday cup.
(c) Diagnostic of $e$-beam position in the Rb chamber. Error bars on the plot are derived from uncertainty of the fit and smaller than the points on the graph.
(d) Fit of recovered Rydberg current vs. Faraday cup current. Error bars on the plot are derived from uncertainty of the fit and smaller than the points on the graph.}
\label{fig:beam_diagnostics}
\end{figure*}

%Our $e$-beam is generated from a Staib Instruments tungsten hairpin source. 
A more detailed view of the experimental arrangements in Fig.~\ref{fig:setup}(d) shows the location of grounded copper gaskets whose role is to attenuate possible electric charge buildup the chamber windows from scattered electrons or photoelectrons produced by the blue laser~\cite{MaOE20,patrick2025imaginginducedsurfacecharge}. The $e$-beam has an energy of 20~keV and variable current from 0~-~40~$\mu$A that is monitored on a Faraday cup. Unfortunately, our apparatus lacks the ability to verify the electron beam current at the location of the laser beam. %Instead we only know the electron gun emission current on the $e$-beam controller and the current reaching the Faraday cup at the end, and there is a large discrepancy between these two values due to e-beam scattering inside the apparatus. 
%This discrepancy means that the e-beam is larger than the Faraday cup and not fully captured, or is clipped somewhere along the beamline.
%An integrated parametric current transformer (IPCT) was placed along the beamline to validate the current reported by the Faraday cup.
%The electric field generated by the e-beam is measured with rubidium atoms excited to a Rydberg state shown in Fig.~\ref{fig:setup} (b).
%We target the 58D Rydberg state because it strikes a balance between a large electric polarizability and strong signal to noise for our limited blue laser power.

However, we can obtain \textit{in situ} information about the $e$-beam position and size in the Rb chamber by measuring beam impact fluorescence (IF)  -- the incoherent electron impact-induced fluorescence in Rb~\cite{DeStefano_APL2024,ImpactFluorescence,CameraBeam}.
In this case all the lasers are blocked, and the Rb fluorescence caused solely by the $e$-beam is recorded with a 10 s exposure time using the same imaging system.
%The e-beam will excite the Rb atoms which fluoresce.
IF offers a secondary \textit{in situ} measurement of the $e$-beam position and size to compare with the Rydberg measurements.

%Atomic measurements take place within a Rb chamber shown in Fig.~\ref{fig:setup} (c) - (d).
%This chamber is a cell that could in principle be put on any beam line.
%Our chamber contains copper gaskets for the purpose of blocking parasitic fields that may build up on the windows of the chamber when the e-beam is turned on and photo-illuminated fields generated by the blue laser on the chamber windows~\cite{MaOE20,patrick2025imaginginducedsurfacecharge}.
%In hindsight, this design could be improved by making the cell longer along to laser propagation axis, therefore leading to a region of more uniform non-zero electric field.

Fig.~\ref{fig:fluorescence-detection} shows an example of recorded fluorescence spectra and resulting $e$-beam profile analysis.
Fig.~\ref{fig:fluorescence-detection}(a) shows three samples of the EIT fluorescence spectra for different positions along the laser beam path [Fig.~\ref{fig:fluorescence-detection}(b)]: (1) at the region of minimal electric field; (2) at the edge of the $e$-beam where its electric field is the highest; (3) near the surface of the copper gasket where  residual charging produces relatively strong electric field. 
%The number in each plot shows where they lie along the laser beam fluorescence position in Fig.~\ref{fig:fluorescence-detection} (b).
To extract the value of the electric field in each point, we follow the procedure similar to the previously published work~\cite{Noah2DFields,patrick2025imaginginducedsurfacecharge}, where the total EIT fluorescence spectra $\mathcal{F}$ is modeled as a combination of three resonances:
\begin{equation}
\label{eq:sum_of_gaussians}
    \mathcal{F} = A\sum_{|m_j|} w_{|m_j|} \text{exp}\left[\frac{-[\Delta_C - \Delta f_{|m_j|}(E)]^2}{2 \gamma_{_{EIT}}^2}\right].
\end{equation}
Here, $w_{|m_j|}$ are amplitudes for each $|m_j|$ level of the Rydberg state empirically set and constant for all fits, $\Delta_C$ is the coupling laser frequency detuning, $\gamma_{_{EIT}}$ is the linewidth of the EIT resonance set constant for all fits, and $\Delta f_{|m_j|}(E)$ is the frequency shift of the Rydberg energy level described by the Stark map from ARC~\cite{Noah2DFields,patrick2025imaginginducedsurfacecharge,ARC}.
In this fit, the only free parameters are the total amplitude of the EIT profile, $A$, and the value of electric field magnitude, $E$.

The reconstructed electric field distribution is shown in Fig.~\ref{fig:fluorescence-detection} (c). In the center, one can clearly see the characteristic two-lobed feature, expected when the laser beams go through the center of the $e$-beam,  as shown in Fig.\ref{fig:setup}(a). 
%For some values of electric fields the EIT lineshape splits, but for smaller fields it only broadens the resonance.
%The electric field either splits or broadens the EIT lineshape.
%When fitting a single curve with a sum of Gaussian peaks, the minimum detectable variation is given by
%\begin{equation}
%    \Delta f_{|m_j|, min} \simeq \frac{\gamma_{_{EIT}}}{SNR \sqrt{n}},
%    \label{eq:peak determination}
%\end{equation}
%where $SNR$ is the signal to noise ratio of the peak and $n$ is the number of data points in the resonance~\cite{NMRBook,Mikhailov:09}.
%For the signal to noise in our experiment, the minimum shift is $\approx$ 0.1 MHz corresponding to an electric field of $\approx$ 0.02 V/cm.
%Below this value, the reconstructed electric field values are not trustworthy.
The position and width of this feature matches well with the IF signal, shown in Fig.~\ref{fig:fluorescence-detection}(d).
The background electric field near the edges is not directly related to the $e$-beam, but caused by residual charging of the cell windows and walls. 
Our ability to measure small values of electric fields is limited by our ability to resolve small relative shifts of the three EIT peaks. We can estimate the minimum detectable shift to be $\Delta f_{|m_j|, min} \simeq \gamma_{_{EIT}}/(SNR \sqrt{n})$, 
where $SNR$ is the signal to noise ratio, and $n$ is the number of data points in the resonance~\cite{NMRBook,Mikhailov:09}. For our current experiment, this minimum shift is $\approx 0.1$~MHz corresponding to an minimum electric field of $E_{min}\approx0.02$~V/cm.

To extract the $e$-beam parameters, we fit the reconstructed curve with an analytical expression for the electric field produced by an electron beam with a uniform Gaussian distribution: 
%The reconstructed electric field curve is fit with an electric field distribution derived 
\begin{equation}
    E_{e-beam} = \frac{I}{2\pi \epsilon_0 v_e} \frac{1}{r}(1-e^{-\frac{r^2}{{\sigma^2}}}).
    \label{eq:e_field_e_beam}
\end{equation}
Here $\sigma$ is the half-width at half maximum of the $e$-beam, $\epsilon_0$ is the permittivity of free space, $r = \sqrt{(z + \Delta z)^2 + y^2}$ is the radial position away from the $e$-beam center, and $\Delta z$ is its displacement along the $z$-direction, $I$ is the $e$-beam current, and $v_e$ is the speed of the electron, proportional to the square root of the beam energy.
The fit also accounts for the finite width of the laser beam and the $y$-deflection from center of the $e$-beam.
The free parameters in this fit are the $e$-beam width $\sigma$, $e$-beam current $I$, and the $z$ and $y$-displacement of the electron and laser beams respectively.
%To fit this curve, we model a truly Gaussian electric field distribution with parameters of vertical and horizontal e-beam displacement, e-beam current, and e-beam FWHM.
%We fit the reconstructed electric field with this function to extract the parameters of the $e$-beam.

The reconstructed $e$-beam cross section is plotted on top of the IF measurement in Fig.~\ref{fig:fluorescence-detection} (d), showing excellent agreement between the two measurement methods.
%
%For the data presented in this letter, the FWHM is fixed and monitored with the IF measurements.
%and optical transition radiation (OTR).
%OTR is a method of monitoring the e-beam where it strikes a copper plate, and the emitted light is monitored upstream of the Rb chamber in the vacuum system.
The $e$-beam width is measured to be 1.07 $\pm$ 0.06~mm with IF method and 1.1 $\pm$ 0.1~mm with Rydberg electric field reconstruction method.
%We find good agreement with the two methods to measure $e$-beam width where the IF has a value of 1.07 $\pm$ 0.06 mm the reconstructed width from the electric field has a value of 1.1 $\pm$ 0.1 mm. 
The value of the width is measured from repeated measurements with a fixed Faraday cup current of 35 $\mu$A.

Further diagnostics of the $e$-beam position and current are shown in Fig.~\ref{fig:beam_diagnostics}.
%To monitor the $e$-beam center, we again use IF measurements. 
For these measurements we either moved the $e$-beam [Fig.~\ref{fig:beam_diagnostics}(a)] or changed its current [Fig.~\ref{fig:beam_diagnostics} (b)] and tracked the variations in the reconstructed electric field distribution. We see the anticipated variation in the fluorescence spectra. For example, if the beam position changes, the two lobes shift by the corresponding amount.  %spectra and electric field shifts along the probe laser fluorescence region for motion of the $e$-beam.
% shows the correlation between the center of the IF in the chamber and the beam center found through the reconstructed electric field fit. 
%We find that these values are highly correlated through a linear fit that shows both methods yield nearly the same center position.
%For the two fits, the position has R${}^2$ = 0.997 and the current measurements have R${}^2$ = 0.827 \red{what is R? and is it a variable?}.
%Our spatial resolution is limited by the pixel size of the camera. 
%Using our calibration one pixel on the imaging plane is $\approx$ 30 $\mu$m. 
%Our current imaging system can be precise to within this range but this can be improved with a better imaging system.
%To determine a minimum detectable resolution of the motion of the beam center, the mean fit error of the electric field is 8 $\mu$m.
To verify the accuracy of our method, we fit the reconstructed electric field with a functional form, as described above. Fig.~\ref{fig:beam_diagnostics}(c) shows excellent agreement between the reconstructed beam position with that measured using IF method. % that can determine the central position with less deviation than our imaging system.
The shown uncertainties for the Rydberg EIT methods are from the error of the fit parameters, and the centroid position is known within 8~$\mu$m.
%The error extracted from the fit shows that the measurements differ within 8 $\mu$m, setting the precision of our measurement.

Changing the emission current seems to also slightly deform and shift the $e$-beam due to electron repulsion or focusing effects that we see in both electric field reconstruction [Fig.~\ref{fig:beam_diagnostics}(b)] and IF measurements. %Fig.~\ref{fig:beam_diagnostics} (b) and (d) shows the comparison of the $e$-beam parameters, obtained using Rydberg EIT method with the available alternatives..
%In Fig.~\ref{fig:beam_diagnostics} (b), from the IF profiles, %changing the current density also 
Both profiles show an asymmetry that may be due to the background field present within the Rb chamber visible near the edges.
% of the fluorescence region in Fig.~\ref{fig:fluorescence-detection} (d).
%This could be due to variations in the background electric field or another mechanism in the focusing of the beam itself.
%This is most likely due to the influence of the large parasitic field within the Rb chamber that is visible in Fig.~\ref{fig:fluorescence-detection} (d) when the beam is blocked.
%
Fig.~\ref{fig:beam_diagnostics}(d) shows clear linear correlation between the recorded Faraday cup current and the current values reconstructed from the electric field. The reconstructed current values are about twice larger than the measured Faraday cup current, possibly due to the $e$-beam clipping somewhere in the beamline before the Faraday cup.
Further verification will require accurate \textit{in situ} $e$-beam current measurements using, e.g., a co-located harp scanner. 
%can be achieved by calibrating the relative change in the electric field against other methods such as an \textit{in situ} harp scanner or IPCT closer to the Rb chamber.
%While it would be nice to have a truly accurate measurement of the current that was only based on electric field measurements and agreed with other methods, the relative change of electric field could be calibrated to be used as a current monitor.
% This is a hear me out, but it could work paragraph I don't know how common this is to put in papers, but might cover us
%The reconstructed current could be off of the true value for a multitude of reasons.
%
Since our current approach is only sensitive to the absolute value of the electric field, our current fit model does not take into account the direction of the electric field, and its influence on the shifts of the different $|m_j|$ levels of the Rydberg state. In addition, presence of the background electric field can produce additional systematic error in beam reconstruction due to its unknown direction.  
%For different current values of the $e$-beam, it can slightly distort the beam shape as shown in the $e$-beam fluorescence measurements shown in Fig.~\ref{fig:beam_diagnostics} (c).
%It is also unclear how the background field affects the $e$-beam or its generated field.
%Measurements of the direction of the electric field produced by the $e$-beam and 
More accurate current measurements may require improvement of the Rb vapor chamber design to further reduce the parasitic charging.

%\section{Conclusion}

In conclusion, we apply Rydberg fluorescence-based detection to measure spatially varying dc electric fields  produced by an $e$-beam and to reconstruct the center-of-mass beam position to within 8~$\mu$m, determine the beam width to within 100~$\mu$m, and measure the beam current in a simultaneous measurement.
%Our apparatus does this measurement slowly and has a parasitic background field that somewhat distorts the measurement, but these are technical limitations that can be easily overcome.
%One e-beam characterization that we struggle in is finding the current, but we present an argument for why this is difficult, and hard to test in our current apparatus.
%While in it's current state this measurement technique is slow, this is a technical limitation with our equipment and can in principle be done much faster with another camera or stronger lasers.
We expect this technique to be useful for diagnostics of  charged particle beams of any energy, and for diagnostics of charged particle in general, such as low-density plasmas ~\cite{Mordjick_BRN2024}.
Further diagnostics such as full beam cross-sectional profiling can be achieved if a sheet of light is used~\cite{Noah2DFields}, and the speed can be improved using faster camera.
%direction of the field can be inferred if a more robust Rb chamber is constructed.

\section{Acknowledgments}
The authors thank Saeed Pegahan and Ziqi Niu for help with experiments, and Chris Holloway's group at NIST for enlightening conversations.

This work is supported by U.S. DOE Contract DE-SC0024621 and DE-AC05-06OR23177, NSF award 2326736 and Jefferson Lab LDRD program.


%\section{Supplementary}

% \subsection{Minimum Detectable Electric Field}
% The electric field either splits or broadens the EIT lineshape.
% When fitting a single curve with a sum of Gaussian peaks, the minimum detectable variation is given by
% \begin{equation}
%     \Delta f_{|m_j|, min} \simeq \frac{\gamma_{_{EIT}}}{SNR \sqrt{n}},
%     \label{eq:peak determination}
% \end{equation}
% where $SNR$ is the signal to noise ratio of the peak and $n$ is the number of data points in the resonance~\cite{NMRBook,Mikhailov:09}.
% For the signal to noise in our experiment, the minimum shift is $\approx$ 0.1 MHz corresponding to an electric field of $\approx$ 0.02 V/cm.

%\subsection{Alternative Beam Tracker Method without Profiel Reconstruction}
%As an alternative, a faster measurement can be devised as follows.
%If we look at the signal-to-noise ratio and measure the slope of the line near the center of the feature in the reconstructed electric field, we find that $\Delta z_{min} = E_{noise} \times \frac{d \Delta z}{d E} \approx$ 50 $\mu$m.
%This means that the fit error \red{is more certain measurement--what does this mean?}, but in principle, faster data collection can be done with slightly worse spatial resolution.

\end{document}